\documentclass[12pt]{article}
\pdfoutput=1

\usepackage{subcaption} 

\usepackage[utf8]{inputenc}
\usepackage[tbtags]{amsmath} 				
\usepackage{amssymb, mathrsfs}			
\usepackage{physics}
\usepackage{graphicx}			
\usepackage{tensor}				
\usepackage[makeroom]{cancel}	
\usepackage{bm}								
\usepackage[
      colorlinks=true,
      linkcolor=blue,
      urlcolor=blue,
      filecolor=black,
      citecolor=red,
      pdfstartview=FitV,
      pdftitle={},
      pdfauthor={Michael Gutperle, Nicholas Klein, Dikshant Rathore},
      bookmarksopen=true
      ]{hyperref}
\usepackage{calc}
\usepackage{sectsty}
\usepackage{dsfont}
\usepackage{cite}

\newlength\myheight
\newlength\mydepth
\settototalheight\myheight{Xygp}
\newcommand{\AdS}{\text{AdS}}

\newcommand{\cN}{\mathcal N}

\sectionfont{\large}

\renewcommand{\thefootnote}{\fnsymbol{footnote}}
\renewcommand{\thanks}[1]{\footnote{#1}}
\newcommand{\starttext}{
\setcounter{footnote}{0}
\renewcommand{\thefootnote}{\arabic{footnote}}}
\renewcommand{\epsilon}{\varepsilon}	

\numberwithin{equation}{section} 		

\numberwithin{equation}{section}

\long\def\symbolfootnote[#1]#2{\begingroup%
\def\thefootnote{\fnsymbol{footnote}}\footnote[#1]{#2}\endgroup}

\begin{document}
\setlength{\baselineskip}{16pt}

\starttext
\setcounter{footnote}{0}

\begin{flushright}
\today
\end{flushright}

\bigskip

\begin{center}

{\Large \bf  Holographic 6d co-dimension 2 defect solutions in M-theory }

\vskip 0.4in

{\large Michael Gutperle, Nicholas Klein,   and Dikshant Rathore }

\vskip 0.2in

{\sl Mani L.~Bhaumik Institute for Theoretical Physics}\\
{\sl Department of Physics and Astronomy }\\
{\sl University of California, Los Angeles, CA 90095, USA}

\end{center}
 
\bigskip
 
\begin{abstract}
\setlength{\baselineskip}{16pt}

We consider the uplift of co-dimension two defect solutions of seven dimensional gauged supergravity to eleven dimensions, previously found by two of the authors.   The uplifted solutions are expressed as Lin-Lunin-Maldacena solutions and an infinite family of regular solutions describing holographic defects is found using the electrostatic formulation of LLM solutions.

\end{abstract}

\setcounter{equation}{0}
\setcounter{footnote}{0}

\newpage

\section{Introduction}

In this paper, we will discuss the uplift of solutions of seven dimensional gauged supergravity  of \cite{Gutperle:2022pgw} to eleven dimensions.  These solutions describe  holographic duals of co-dimension two defects in six dimensional SCFTs. The defects preserve  four dimensional conformal symmetry as well as transverse rotational symmetry.  

There are several approaches to constructing holographic duals  of  such defects. First, probe  branes can be placed inside the AdS vacuum of the ten or eleven dimensional theory \cite{Karch:2000gx,DeWolfe:2001pq}. 
The resulting embedding  realizes the unbroken symmetries of the defect,  which is localized at the intersection of the probe brane and the boundary of  AdS. Second, one can construct solutions of the ten or eleven dimensional supergravity with the ansatz of a warped product of AdS and sphere factors which realize the defect symmetries and solve the supergravity Killing spinor equations to obtain a half-BPS solution.  The second approach is generally quite involved and leads to  ``bubbling" solutions, see e.g. \cite{Lin:2004nb,Lin:2005nh,Lunin:2006xr,Gomis:2007fi,DHoker:2007mci,DHoker:2007zhm,DHoker:2007hhe,DHoker:2008lup,DHoker:2008rje}.

A more pedestrian approach is to consider a truncation of the ten or eleven dimensional theory to a lower dimensional gauged supergravity and construct solutions there. Generally, the ansatz and the BPS conditions following from the vanishing of the supersymmetry transformations are easier to solve in the lower dimensions than in higher dimensions. In many cases, such a lower dimensional solution can then be uplifted to the ten or eleven-dimensional supergravity and given a microscopic understanding by relating it to bubbling solutions.

In this paper we will perform an uplift of the solutions found in \cite{Gutperle:2022pgw} and embed it  into a class of  LLM solutions  of M-theory \cite{Lin:2004nb,Gaiotto:2009gz}. The seven dimensional solutions are constructed by warping $AdS_5\times S^1$ over an interval with $U(1)\times U(1)$ gauge fields   along the circle direction. They are related to hyperbolic (topological) black hole solutions by a double analytic continuation.  These solutions have been used recently to construct spindle compactifications \cite{Ferrero:2020laf,Ferrero:2021etw,Ferrero:2021wvk,Faedo:2021nub,Couzens:2021rlk,Hosseini:2021fge,Boido:2021szx,Giri:2021xta,Suh:2021ifj}\footnote{The hyperbolic black holes where also used to calculate charged R\'enyi entropies in holography, see e.g. \cite{Belin:2013uta,Nishioka:2014mwa,Hosseini:2019and}.}. In this case, the warping coordinate takes values on a finite interval and the $S^1$ closes off at the ends of the interval. One ends up with a topological two sphere with  two conical deficits $2\pi(1-{1\over n_{n/s}}), n_{n/s}\in \mathbb{Z}$  at the north and south pole of the sphere. In our solution, the warping coordinate is a semi-infinite interval and the solution describes a co-dimension two defect in a six-dimensional SCFT. We note that the bulk gauge fields are dual to  conserved currents in the CFT and the supergravity solution corresponds to turning on a source for these currents in the plane transverse to the defect.  This means that these defects are twist/disorder defects where fields charged under these currents are picking up a phase when going around the defect. We list some examples of  holographic co-dimension two defect solutions in supergravities in various dimensions \cite{Gutperle:2018fea,Gutperle:2019dqf,Chen:2020mtv}.

The structure of the present paper is as follows:  In  section \ref{sec2}, we review the defect solution of 
\cite{Gutperle:2022pgw}, in particular, the conditions for a completely non-singular solution with two non-vanishing gauge fields  and a solution with a conical singularity in the bulk with only one gauge field turned on.  In section \ref{sec3} we use the formulas from \cite{Cvetic:1999xp} to lift the seven dimensional solution to eleven dimensions and investigate the nature of the conical singularity of the one charge solution. In section \ref{sec4} we bring the uplifted one charge solution into canonical LLM form. Since our solution has an extra rotational symmetry the LLM solution can be described by an electrostatic potential by a change of variables and we determine the line charge distribution  associated with the one charge solution. This allows us to identify the conical singularity with a ``regular puncture" which was previously discussed in the context of the LLM construction of duals of $d=4, \; \cN=2$ SCFTs by Gaiotto and Maldacena \cite{Gaiotto:2009gz}. In addition, it allows us to construct generalized solutions with more complicated line charge distributions, some of which are completely regular. We calculate  holographic observables namely the on-shell action and the vacuum-subtracted defect central charge. In  appendix \ref{appendixa} we construct a simple example for a co-dimension two defect in a $d=6, \; \cN=(2,0)$ SCFT using the six dimensional free tensor multiplet.

\section{Seven dimensional solution}
\label{sec2}

The seven dimensional supergravity theory   is a truncation of the maximal $d=7$ $SO(5)$ gauged supergravity, where we keep  two scalars and two $U(1)$ gauge fields. The theory is defined by the Lagrangian \cite{Cvetic:1999xp}
\begin{align}
S&= \int d^7x \sqrt{-g} \Big( R- {1\over 2} \sum_{i=1}^2 \partial_\mu \varphi_i \partial^\mu \varphi_i - g^2 V-{1\over 4} \sum_{i=1}^2e^{\vec{a}_i \vec{\phi}} F_{(i)}^2\Big),
\end{align}
where we use
\begin{align}
\vec{\alpha}_1&= (\sqrt{2},\sqrt{2\over 5}) , \quad \vec{\alpha}_2= (-\sqrt{2},\sqrt{2\over 5})
\end{align}
to define
\begin{align}
X_1&= e^{-{1\over 2} \vec{\alpha}_1 \vec{\varphi}}, \quad X_2= e^{-{1\over 2} \vec{\alpha}_2 \vec{\varphi}}, \quad  X_0 = (X_1 X_2)^{-2},
\end{align}
 and the potential $V$ can  the be expressed as
\begin{align}
V= - 4X_1 X_2 - 2 X_0 X_1 -2 X_0 X_2+{1\over 2} X_0^2.
\end{align}

We consider the following solution of the gauged supergravity  which  can be obtained by a double analytic continuation of charged black hole solutions \cite{Cvetic:1999xp,Cvetic:1999ne,Liu:1999ai}. These have been used to describe  M5 branes wrapped on spindles \cite{Ferrero:2021wvk}, duals of $d=4, \;\cN=2$ Argyres-Douglass theories \cite{Bah:2021mzw,Bah:2021hei},
 and co-dimension 2 defects  \cite{Gutperle:2022pgw} in this theory.
\begin{align}\label{metseven}
ds_7^2&=  \Big( y P(y) \Big)^{1\over 5} ds_{AdS_5}^2 + { y^{6\over 5}   P(y)^{1\over 5} \over 4 Q(y)} dy^2 +  { y^{1\over 5} Q(y) \over P(y)^{4\over5}} dz^2, \nonumber\\
P(y) &= h_1(y) h_2(y), \quad Q(y)= -y^3+ \mu y^2 +{1\over 4} g^2 h_1(y) h_2(y).
\end{align}
The functions $h_i, i=1,2$ are given by
\begin{align}
h_1&= y^2+q_1, \quad h_2= y^2 + q_2.
\end{align}
The scalar fields are expressed in terms of $h_i$ as follows
\begin{align}
X_1&= y^{2\over 5} {h_2(y)^{2\over 5} \over h_1(y)^{3\over 5}},\quad X_2= y^{2\over 5} {h_1(y)^{2\over 5} \over h_2(y)^{3\over 5}},
\end{align}
and the two $U(1)$ gauge fields are given by
\begin{align}
A_1 &=  {\sqrt {1-{\mu\over q_1}} q_1 \over h_1(y)} dz+ a_1 dz , \quad A_2 = {\sqrt {1-{\mu\over q_2}} q_2 \over h_2(y)}  dz+ a_2 dz.
\end{align}
The constant $\mu$ is an extremality parameter and supersymmetric solutions are obtained by setting $\mu=0$. A solution with both $q_1,q_2$ nonzero will preserve one-quarter of the supersymmetry and, as we shall review in the next section, completely nonsingular solutions are possible.  Setting $q_2=0$ produces a solution that preserves half the supersymmetry of the seven dimensional gauged supergravity but such a solution suffers from conical singularities. For the gauge field to be non-singular at the location $y=y_+$, where the space closes off, we have to choose $a_1$ and $a_2$ such that
\begin{align}\label{regularas}
    A_1(y_+)= A_2(y_+)=0.
\end{align}

In the following we set the coupling $g=2$. As discussed below, this implies that the asymptotic boundary $AdS_5\times S^1$ is conformal to $\mathbb{R}^{1,5}$ without a conical deficit provided $z$ has standard periodicity $2\pi.$

\begin{figure}[h]
\centering
\includegraphics[scale=.8]{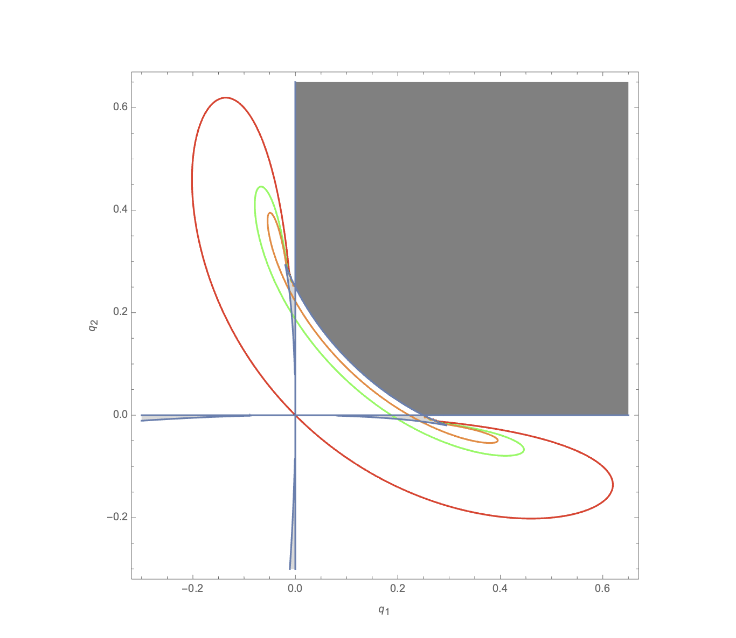}
\caption{\textbf{Regular two charge solutions.} Allowed charges for different values of conical deficits: $n=1$ (red) is completely regular. $n=2$ (green), $n=3$ (orange) correspond to the first two half-spindle solutions. The dark grey portion is the disallowed region where $Q(y)$ has no real zeros. }
\label{Fig1}
\end{figure}

\subsection{Regular two charge solution}
\label{sec2:twocharge}

The case of completely regular solutions was analyzed in \cite{Gutperle:2022pgw}. These solutions were constructed by allowing the warping coordinate $y$ to take values in the semi-infinite interval $[y_+,\infty]$ where $y_+$ is the largest zero of $Q(y)$ defined in (\ref{metseven}). The existence of such a positive $y_+,$ which produces no double zero and a regular metric everywhere, is guaranteed as long as we place conditions (discussed in \cite{Gutperle:2022pgw}) on the signs of the charges $q_1,q_2$ as well as the discriminant of the polynomial $Q(y)$. 

This interval produces a non-compact space and therefore, unlike in the spindle construction, we approach the asymptotic $AdS_7$ region as $y\to\infty$. In this limit the metric (\ref{metseven}) takes the form 
\begin{align}
\lim_{y\to\infty} ds^2_7 &= yds^2_{AdS_5} + ydz^2 +{1\over 4y^2}dy^2 + ...\nonumber\\
&= {d\rho ^2 \over 4\rho^2} + {1\over \rho} \left(ds^2_{AdS_5} +dz^2\right)+ ...,
\end{align}
where we make the change of coordinates $y = 1/\rho$ and the dots denote subleading terms. Note that the boundary of this space is of the form $AdS_5\times S^1$ which is conformal to $\mathbb{R}^{1,5}$ with no conical defect as long as the coordinate $z$ parameterizing the $S^1$ has periodicity $2\pi$. 

Having fixed the periodicity of $z,$ we can look at the metric in the region $y\to y_+$. Letting $y = y_+ + \rho$, we have that $Q(y)\approx Q'(y_+) \rho$ and $P(y)\approx P(y_+) = y_+^3$ so that the metric (\ref{metseven}) takes the form

\begin{align}
\left(yP(y)\right)^{1/5}\left( {y\over 4Q(y)} dy^2 + {Q(y) \over P(y)} dz^2 \right)\approx {y^{9/5} \over Q'(y_+)}\left(dr^2 + \left({Q'(y_+) \over y_+^2}\right)r^2dz^2\right),
\end{align}
where we define the new radial coordinate $r = \rho^{1/2}$. Notice that at $r=0$ ($y=y_+$) the $z$-circle shrinks to zero size and the space closes off. At this location, we may fix the values of $q_1,q_2$ such that we either have a regular solution or a $\mathbb{R}^2/\mathbb{Z}_k$ singularity by setting:

\begin{align}\label{Qconstraint}
{Q'(y_+) \over y_+^2} = {1\over k}.
\end{align}
The values $k>1$ give the metric  with deficit angle $2\pi(1-1/k)$ at $y=y_+$. Using the explicit form of the function $Q(y)$, we can express the constraint (\ref{Qconstraint}) as

\begin{align}
y_+\left(4y_+^2 - (3+1/k)y_+ + 2(q_1+q_2)\right) = 0.
\end{align}
Note that the root $y_+$ itself depends on the charges $q_1,q_2$ however we can clearly see that the above condition will constrain them to lie along a different one dimensional curve for each choice of $k.$ In figure \ref{Fig1}, we have plotted the first three of these families of solutions in the $q_1,q_2$-plane. 

\subsection{One charge solution}
\label{sec2:onecharge}

The solution with two nonzero charges is quarter BPS, i.e. preserves eight of the original thirty-two supersymmetries of the $d=7$ gauged supergravity. Our goal is to obtain solutions which fit into the LLM solutions in 11 dimensions, which preserve sixteen supersymmetries. We will have to set one of the two charges to zero in order to produce a half BPS solution. In the following we will set   $q_2$ to zero. The metric components of (\ref{metseven}) in the $y$ and $z$ direction become (recall that we have set $g=2$)
\begin{align}
    ds_7^2 = {(y^2+q_1)^{1\over 5} \over   4y^{2\over 5} \big( y^2+q_1- y\big)}dy^2+ {y^{3/5} \big( y^2+q_1- y\big) \over  (y^2+q_1)^{4\over 5}} dz^2 +\cdots.
\end{align}
The larger zero of $y$ is located at
\begin{align}\label{ycloc}
    y_c= {1\over2} \big(1+\sqrt{1- 4 q_1}\big).
\end{align}
With the following change of variable
\begin{align}
    y=  y_c +{1\over 4}r^2,
\end{align}
the metric near  $y\sim y_c$, i.e. $r\sim 0$ behaves as follows
\begin{align}
    ds^2 \sim  {1\over 2^{9\over 5} \sqrt{1-4 q_1}(1+\sqrt{1- 4 q_1} )^{1\over 5}} \Big( dr^2 + (1-4 q_1) r^2 dz^2\Big) +\cdots.
\end{align}
Consequently,  for nonzero $q_1$ there is a  conical singularity  in  the bulk of the  spacetime, whereas $q_1=0$ corresponds to the $AdS_7$ vacuum.  For a $\mathbb{R}^2/\mathbb{Z}_k$ conical singularity  with deficit  $2\pi(1-{1\over k})$,  the charge $q_1$  is given by
 \begin{align}\label{q1val}
     {1\over k} = \sqrt{1-4 q_1}.
 \end{align}
 In seven dimensions a conical singularity in the bulk is problematic. In some cases uplifting a singular  solution of lower dimensional supergravity to ten or eleven dimensions leads to a non-singular solution, in other cases the solution may have a well defined interpretation in terms of branes. 

\section{Uplift to eleven dimensions}
\label{sec3}

A solution of seven dimensional gauged supergravity can be uplifted to eleven dimensional supergravity \cite{Cvetic:1999xp},  the metric and the four-form antisymmetric tensor field strength take the following form 
\begin{align}\label{meteleven}
ds_{11}^2 &= \Omega^{1\over 3} ds_7^2 +  {1\over g^2 \Omega^{2\over 3}}\Big\{ {d\mu_0^2\over X_0}+ \sum_{i=1}^2 {1\over X_i} \Big(d\mu_i^2+\mu_i^2(d\phi_i+ g A_i)^2\Big)\Big\},\nonumber\\
*_{11}F_4 &= 2g \sum_{\alpha=0}^2 \Big(X_\alpha^2 \mu_\alpha^2 -\Omega X_\alpha\Big) \epsilon_7+ g \Omega X_0 \epsilon_7 + {1\over 2g} \sum_{\alpha=0}^2 {*_7 d\ln X_\alpha}\wedge d(\mu_\alpha^2)\\
& +{1\over 2 g^2} \sum_{i=1}^2 {1\over X_i^2} d(\mu_i^2)\wedge (d\phi_i+ g A_i)\wedge *_7 F_i, \nonumber
\end{align}
where $F_i= dA_i$ and $*_7$ is the Hodge dual with respect to the seven dimensional metric (\ref{metseven}) and $*_{11}$ the Hodge dual with respect to the eleven dimensional metric (\ref{meteleven}). $\phi_i, i=1,2$ are two angular coordinates with period $2\pi$ and  the variables $\mu_\alpha, \alpha=0,1,2$ parametrize a two sphere
\begin{align}
\mu_0 ^2+\mu_1^2+\mu_2^2=1
\end{align}
and the warp factor  $\Omega$ is given by
\begin{align}
    \Omega=X_0 \mu_0^2+ X_1 \mu_1^2+ X_2 \mu_2^2.
\end{align}
We will parameterize the $\mu_i$ in the following way
\begin{align}\label{alpar}
 \mu_0= \sin\alpha \cos\theta, \quad \mu_1 =\sin\theta, \quad \mu_2= \cos\alpha\cos\theta.  
\end{align}

\subsection{Two charge solution}
With our $\mu_i$ parameterization, the warp factor $\Omega$ becomes
\begin{align}
    \Omega= {(y^2+q_1)^{2\over 5} (y^2+ q_2 \sin^2\alpha) \cos^2\theta \over y^{8\over 5} (y^2+q_2)^{3\over 5}} + {y^{2\over 5} (y^2+q_2)^{2\over 5} \sin^2\theta \over (y^2+q_1)^{3\over 5}}.
\end{align}
As discussed in section  \ref{sec2:twocharge}, $y^2+q_1>0$ and $y_2+q_2>0$ for $y\geq y_c$ for the solutions which satisfy the regularity conditions. Hence, if the seven dimensional metric is regular then the eleven dimensional metric is also regular and describes a quarter-BPS co-dimension two defect in M-theory.

\subsection{One charge solution}

The uplift of the  $q_2=0$  solution given in section \ref{sec2:onecharge} 
and the eleven dimensional metric for the defect solution takes the following form
\begin{align}\label{metb}
ds_{11}^2&= \kappa^{2\over 3} \Big\{ y^{1\over 3} ( y^2+ q_1\cos^2\theta)^{1\over 3} ds_{AdS_5}^2+  {y^{4\over3}\cos^2 \theta \over 4  (y^2+ q_1\cos^2\theta)^{2\over 3} }ds_{S_2}^2+ {(y^2+ q_1\cos^2\theta)^{1\over 3}\over 4 y^{2\over 3}}d\theta^2\nonumber\\
&+{(y^2+q_1\cos^2\theta )^{1\over 3}\over 4 y^{2\over 3}  (y^2- y+q_1)}dy^2+ {y^{1\over 3}(y^2+q_1\cos^2\theta)^{1\over 3} \big(y^2- y+ q_1\big) \over  (y^2+q_1)} dz^2 \\
&+{(y^2+q_1) \sin^2\theta \over 4 y^{2\over 3} (y^2 +q_1\cos^2\theta)^{2\over 3}} (d\phi_1 + {2 q_1\over y^2 +q_1}dz+ 2 a_1 dz)^2 \Big\}, \nonumber
\end{align}
where we used  the  parameterization (\ref{alpar}) for $\mu_\alpha,\alpha=0,1,2$. The coordinates $\alpha$ and $\phi_2$ will parameterize the round two sphere
\begin{align}
ds_{S_2}^2= d\alpha +\sin^2 \alpha \; d\phi_2^2.
\end{align}
The uplifted metric therefore geometrically realizes an $SU(2)$ symmetry, which will be interpreted as an R-symmetry from the perspective of the four dimensional $\cN=2$ defect theory.
Using the uplift  formula (\ref{meteleven}), one obtains for the four form 
\begin{align}\label{fourform1}
F_4&=   \kappa \Big\{  vol(S_2)\wedge(  f_{\phi_1}  d\phi_1 +f_z dz)\wedge d\theta + vol(S_2) \wedge (  g_{\phi_1}  d\phi_1 +g_z dz)  \wedge dy\Big\}
\end{align}
with
\begin{align}
f_{\phi_1} &= {(y^2+q_1)(3y^2+ q_1 \cos^2\theta) \cos^2 \theta\sin\theta\over 8 (y^2+q_1 \cos^2\theta)^2},\nonumber \\
f_z&= { \big( q_1 + a_1 (y^2+q_1) \big) (3y^2+ q_1 \cos^2\theta)\cos^2\theta\sin\theta\over 4 (y^2+q_1 \cos^2\theta)^2},\nonumber\\
g_{\phi_1}&=  {q_1 y \cos^3 \theta \sin^2\theta\over 4 (y^2+q_1 \cos^2\theta)^2}, \\ 
g_z&= {q_1 y (1+ a_1 \sin^2\theta) \cos^3 \theta \over 2(y^2+q_1 \cos^2\theta)^2}. \nonumber
\end{align}

Note that in contrast to  solutions where $y$ takes values on a compact interval, in our case the region $y\to \infty$  is part of the spacetime and corresponds  to the asymptotic $AdS_7\times S^4$  region. In this limit, the metric  and the four form behave as follows 
\begin{align}\label{asymptoticAdS7metric}
    ds^2 & \sim \kappa^{2\over 3} \Big( y ds_{AdS_5} + {1\over 4} \cos^2\theta ds_{S^2}^2+ {1\over 4} d\theta^2+{1\over 4 y^2} dy^2+ y dz^2 + {1\over 4} \sin^2 \theta (d\phi_1+ 2 a_1 dz)^2+ \mathcal{O} ({1/ y}),\nonumber\\
    F_4&\sim   \kappa\;   {3\over 8}\cos^2\theta \sin\theta  \; { vol}(S^2)\wedge ( d\phi_1  + 2 a_1 dz) \wedge d\theta + \mathcal{O} ({1/y}).
\end{align} 
The angular coordinates $z, \phi_1$ have period $2\pi$. We can define a new angular coordinate $\tilde \phi= \phi_1+ 2 a_1 z$, which has standard period $2\pi$ for  $a_1= k/2, k\in \mathbb{Z}$.
The flux of the four form on the $S^4$ is given by
\begin{align}
\int F_4&= \kappa {3\over 8} \int_{S^2} vol(S^2) \int_0^\pi  d\theta   \cos^2\theta \sin\theta \int_0^{2\pi} d\tilde\phi\nonumber\\
&= 2\pi^2 \kappa = {16\over g^3} \pi^2 \kappa
 \end{align}
 where we  restored the gauge coupling $g.$
The condition for charge quantization for the four form $F_4$ in M-theory   is given by
\begin{align}
{1\over (2\pi)^3 \ell_p^3} \int F_4=N, \quad N\in \mathbb{Z},
\end{align}
where $N$ can be interpreted as  the number of fivebranes leading to the  $AdS_7\times S^4$ vacuum in the near horizon limit and hence, the constant $\kappa$ in the uplift formula is
\begin{align}
\kappa= {\pi \over 2 } g^3 N\;  \ell_p^3.
 \end{align}
 Recall that the seven dimensional metric  for the one charge solution has a conical singularity at $y=y_c$ (\ref{ycloc}). Defining $y= y_c + r^2$ and expanding around $r=0,$ the eleven dimensional metric takes the following form
 \begin{align}
 ds^2&\sim { (y_c^2+ q_1\cos^2\theta)^{1\over 3}\over y_c^{2\over 3}}\left\{  y_c ds_{AdS_5}^2  + {y_c^2 \cos^2\theta \over 4 (y_c^2+ q_1\cos^2\theta)} ds_{S^2}^2 +  {d\theta^2\over 4}+ {dr^2 \over \sqrt{1-4q_1}} \right. \nonumber\\
&  \left.  + \sqrt{1-4q_1}   r^2 dz^2 +  {\sqrt{1-4q_1} \sin^2\theta\over 1+  \sqrt{1-4q_1}-2 q_1 \sin^2\theta} \Big( d\phi_1+  (1-\sqrt{1-4q_1} +2 a_1) dz \Big)^2 \right\}+ \mathcal{O}(r^2)
 \end{align}
 There are three potential conical singularities in the  $\theta, z,r,\phi_1$ part of the metric. At $\theta=\pi/2$ the two sphere shrinks to zero size in a smooth way, and at $r=0$ there is a $\mathbb{R}^2/\mathbb{Z}_k$ conical singularity if $1/k=\sqrt{1-4q_1}$ which is inherited from the seven dimensional metric. At $\theta=0$ we can define a new angular variable 
 \begin{align}
 \hat \phi = \phi_1 +\big( 1+ 2 a_1-{1\over k} \big) z.
 \end{align}
 As argued above  from the regularity in the asymptotic $AdS_7\times S^4$ limit, $2a_1$ is an integer and both $\phi_1$ and $z$ have period $2\pi$. Hence the new angular variable $\hat \phi$  has period $2\pi/n$ and the metric displays a $\mathbb{R}^4/\mathbb{Z}_k$ singularity near the point  $r=0, \theta=0$.

\section{Lin-Lunin-Maldacena solutions}
\label{sec4}

The M-theory LLM solutions \cite{Lin:2004nb} are examples of ``bubbling" supergravity solutions which holographically are the  deformation of the $d=6,\; \cN=2$ SCFT by half-BPS states of dimension $\Delta\sim N^2$. In the same paper a double analytic continuation related these solutions to  a  general solution of eleven dimensional supergravity with $SO(2,4)\times SU(2)\times U(1)$ symmetry. These solutions have been used to find  holographic  duals \cite{Gaiotto:2009gz}  of  a large class of $d=4,\; \cN=2$ SCFTs constructed in \cite{Gaiotto:2009we}.  The goal of the present section is to show that our uplifted solution can be written  in the LLM  form. We briefly review the salient features of the LLM solution \cite{Gaiotto:2009gz}. The metric is given by an $AdS_5\times S^2$ warped over a four dimensional space, which is a $U(1)$ fibration over a three dimensional base space spanned by coordinates $\xi, x_1,x_2$

\begin{align}
ds_{11,LLM}^2&=\kappa_{11}^{2\over 3} {e^{2\lambda} } \Big\{ 4 ds_{AdS_5}^2+ {\xi^2e^{-6 \lambda} } ds_{S^2}^2+ {4\over1- \xi \partial_\xi D} (d\chi-{1\over 2} v_i dx^i)^2\nonumber\\
&\quad  \quad \quad-{\partial_\xi D \over \xi} \big(d\xi^2+ e^D(dx_1^2+dx_2^2)\big)\Big\}.
\end{align}
The four form field strength takes the following form
\begin{align}\label{fourform2}
F_4&=2 \kappa_{11}  {\rm vol}(S_2)\wedge\Big(d\chi+ v)\wedge d(\xi^3 e^{-6\lambda})+(\xi-\xi^3 e^{-6\lambda}) dv -{1\over 2} \partial_\xi e^D dx_1\wedge dx_2\Big).
\end{align}

  The dimensionful quantity $\kappa_{11}={\pi\over 2} \ell_p^3$ is the standard choice, note that our $\kappa$ has both $N$ and $g$ in it, this way we have to absorb the charges to $D$ which makes the comparison easier to \cite{Gaiotto:2009gz}. Therefore, we identify $\kappa = g^3 N \kappa_{11}$.

The solution is completely determined in terms of a single function $D(\xi,x_1,x_2)$ 
\begin{align}
e^{-6\lambda} = {-\partial_\xi D \over \xi(1-\xi \partial_\xi D)}, \quad \quad dv= \sum_i{v_i} dx^i, \quad  v_1=-\partial_{x_2} D, \quad v_2 = \partial_{x_1} D.
\end{align}
The function $D(\xi,x_1,x_2)$ satisfies the partial differential equation of  Toda type
\begin{align}\label{todaa}
\big(\partial_{x_1}^2 + \partial_{x_2}^2\big) D + \partial_\xi^2 e^{D}=0.
\end{align}
Our goal is to find the LLM form of our uplifted solution (\ref{metb}). We note that our solution has an additional rotational symmetry in the $x_1,x_2$ plane which allows us to write the metric as
 \begin{align}\label{llmb}
ds_{11,LLM}^2&= \kappa_{11}^{2\over 3} {e^{2\lambda} } \Big\{ 4 ds_{AdS_5}^2+ {\xi^2e^{-6 \lambda} } ds_{S^2}^2+ {4\over1- \xi \partial_\xi D} (d\chi-{\rho\over 2} \partial_\rho D \; d\beta )^2\nonumber\\
&\quad \quad \quad -{\partial_\xi D\xi} \big(d\xi^2+ e^D(d\rho^2 +\rho^2 d\beta^2)\big)\Big\}.
\end{align}
As we will review in section \ref{sec4:u(1)}, this additional symmetry allows for a reformulation in terms of an electrostatic problem \cite{Gaiotto:2009gz,Donos:2010va,Reid-Edwards:2010vpm,Aharony:2012tz,Petropoulos:2013vya} which replaces the Toda equation with a linear Laplace equation.

\subsection{Map to LLM}
In order to find the map of the metric (\ref{metb}) to an LLM form (\ref{llmb}), we note that the metric (\ref{metb}) depends on the two coordinates $y,\theta$ while  the LLM metric with the additional $U(1)$ isometry also depends on two coordinates $\xi, \rho$.  In addition there are two angular coordinates $\phi,z$ which have to be related to $\chi, \beta$.  

By comparing the $AdS_5$ and $S^2$ parts of the two metrics, we can  determine the radial coordinate $\xi$ in terms of $y,\theta$ as well as  an  expression for $\lambda$ in  (\ref{llmb})
\begin{align}\label{varmap}
\xi=N  y \cos \theta,  \quad \quad e^{6\lambda}= N^2\; y(y^2 + q_1 \cos^2\theta),
\end{align}
and we can choose an ansatz for the  second radial coordinate $\rho$
\begin{align}\label{varans}
\rho = \sin \theta\; g(y)
\end{align}
for some function $g(y)$.
Using these relations, the $g_{\xi\xi}, g_{\rho\rho} $  and the $g_{\xi\rho}$ components of  (\ref{llmb}) can be expressed in terms of the $y, \theta$ coordinates and be matched to the uplifted metric (\ref{metb}). This gives us a differential equation for the function $g(y)$
\begin{align}
{d\over dy} \ln g(y)   =  {y \over y^2-y+q_1},
\end{align}
 which can be integrated to obtain
\begin{align}
g(y)&=  \Big(y-{1\over 2} (1+ \sqrt{1-4q_1}) \Big)^{{1\over 2}\big( 1+{1\over \sqrt{1- 4q_1}}\big)}  \Big(y-{1\over 2} (1- \sqrt{1-4q_1}) \Big)^{{1\over 2}\big( 1-{1\over \sqrt{1- 4q_1}}\big)} 
\end{align}
as well as  an expression for the function $D$ expressed as a function of $y$
\begin{align}\label{dexp}
e^D&= N^2{\big(y^2-y+q_1 \big)  \over  g(y)^2}.
\end{align}
The function $D$ satisfies the Toda equation (\ref{todaa}),  which can be verified using the mapping  (\ref{varmap}).
The mapping is complete by finding the identification of angular variables
\begin{align}
z&= c_1 \chi + c_2 \beta, \quad \quad  \phi= c_3 \chi+ c_4\beta.
\end{align}
Matching the  angular components of the metric gives the following relations for $c_i, i=1,\cdots,4$
\begin{align}\label{cmap1}
c_1= \pm 1, \quad c_2 =0, \quad c_3=\mp 2(1+a_1) , \quad c_4= \mp 1.
\end{align}
To match the metric components, both signs in (\ref{cmap1}) are possible, however, matching the four form components (\ref{fourform1}) and (\ref{fourform2}) selects the upper signs.

For the choice of the upper signs in (\ref{cmap1}),  the relations for the angular variables become
\begin{align}
z&=\chi , \quad  \phi=- \beta -2(1+a_1) \chi,
\end{align}
which means that the periodicity of both sets of  angular variables is $2\pi$.

\subsection{ $U(1)$ symmetric solutions}
\label{sec4:u(1)}

The LLM metric (\ref{llmb}) has an additional $U(1)$ symmetry associated with shifts of the angle $\beta.$ For such geometries, it is possible to find an implicit change of variables that turns the nonlinear Toda equation (\ref{todaa}) into a linear Laplace equation. This idea goes back to the paper by Ward \cite{Ward:1990qt} and has been applied to the LLM solution in \cite{Gaiotto:2009gz,Donos:2010va,Reid-Edwards:2010vpm,Aharony:2012tz,Petropoulos:2013vya
}. Note that in some of these papers the $U(1)$ circle  is compactified  to obtain a type IIA solution from the M-theory one.

We  map the LLM coordinates $\xi,\rho$  to the new ones $r, \eta$ and relate the function $D$ to an electrostatic potential
\begin{align}\label{changev}
\rho^2 e^{D(\xi,\rho)} = r^2, \quad   \xi= r \partial_r V \equiv \dot V, \quad \ln \rho = \partial_\eta V \equiv V'.
\end{align}
The function $V(r,\eta)$ satisfies the Laplace equation in cylindrical coordinates
\begin{align}\label{cyllaplace}
{1\over r} \partial_r ( r \partial r V)+ \partial_\eta^2 V=0.
\end{align}
The four dimensional metric and the three form potential  are given by
\begin{align}\label{U1SymSolution}
ds_{11}^2&= \kappa_{11}^{2\over 3} \left( {\dot V \Delta \over 2 V''}\right)^{1\over 3} \Big\{  4 ds_{AdS_5}^2 + {2 V'' \dot V\over \Delta}ds_{S_2}^2+ {2V''\over \dot V}\big( dr^2 + {2\dot V\over 2 \dot V- \ddot V} r^2 d\chi^2+ d\eta^2\Big) \nonumber\\
  &\quad + {2(2 \dot V-\ddot V)\over \dot V \Delta}\big(d\beta+ {2\dot V \dot V'\over 2 \dot V- \ddot V}d\chi\big)^2\Big\},\nonumber\\
C_3&=2 \kappa_{11} \left( -2 {\dot V^2 V''\over \Delta} d\chi+ \big( {\dot V \dot V'\over \Delta} -\eta\big) d\beta\right)\wedge d\Omega_{S^2}
\end{align}
where $d\Omega_{S^2}$ is the volume form on $S^2$ and $\Delta$ is defined as
\begin{align}
\Delta= (2 \dot V-\ddot V)V'' + (\dot V')^2.
\end{align}
To determine the mapping to electrostatic coordinates we are following appendix C in \cite{Petropoulos:2013vya}. The relation (\ref{changev}) gives $r=r(\xi,\rho)$ and the expression for the other variable $\eta=\eta(\xi,\rho)$ implies the exact differential
\begin{align}\label{detadiff}
    d\eta &= {\partial \eta \over \partial \xi}d\xi + {\partial \eta\over \partial\rho}d\rho = {\rho \over r} \partial_{\rho}r d\xi -{r\over \rho} \partial_{\xi}r d\rho.
    \end{align}
   The electrostatic potential can be obtained from the exact differential
   \begin{align}
    dV &=\left( -{r\over \rho} \partial_\xi r \ln \rho +{\xi\over r} \partial_\rho r\right) d\rho + \left( {\xi\over r}\partial_\xi r + {\rho\over r} \partial_\rho r  \; \ln \rho\right) d\xi.
\end{align}
The boundary condition that the sphere closes at $\xi=0$ implies
\begin{align}\label{bcone}
   \partial_r V\mid _{\eta=0} =0.
\end{align}
The rotational symmetric solution corresponds to a conducting disk at $\eta=0$, which is equivalent to (\ref{bcone}) since $ \partial_r V$ is the electrical field in the $r$ direction which vanishes for a conductor at $\eta=0$. 

The potential V is determined by a line charge $\lambda (\eta)$ localized at $r=0$
\begin{align}\label{lambdadef}
    \lambda(\eta)&= r\partial_r V\mid_{r=0}  = \xi(r=0, \eta).
\end{align}
Hence determining the change of variables gives the line charge density.
The potential can then be obtained via the Green's function
\begin{align} \label{vpot}
    V&=-{1\over 2} \int d\eta' G(r,\eta,\eta')\lambda(\eta')
\end{align}
where the Green's function can be obtained by the method of images (adding a line charge at negative $\eta$)
\begin{align}
    G(r,\eta,\eta')={1\over \sqrt{r^2+(\eta-\eta')^2} }- {1\over \sqrt{r^2+(\eta+\eta')^2} }.
\end{align}
    
\medskip

A set of rules for the charge distributions $\lambda(\eta)$ which leads to regular solutions (or those with only $A_k$ singularities) was found in \cite{Gaiotto:2009gz}. The line charges must be piecewise linear and convex with integer slopes. Furthermore, the slopes can only change at integer values of $\eta$. We will say more about these conditions later, but a final point  that we want to explore in this subsection is the relationship between the intercepts of these line segments and the flux of the four form field strength $F_4$.

\begin{figure}[t]
\centerline{\includegraphics[scale=.25]{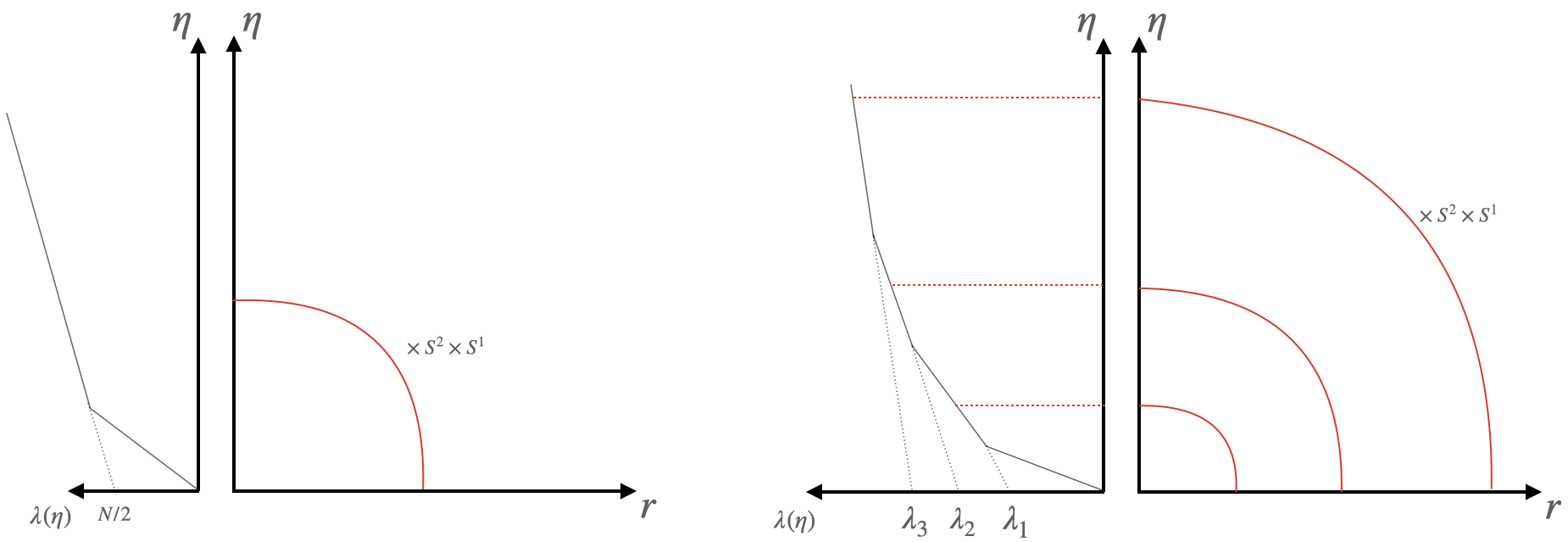}}
\caption{\textbf{Left:} An arc in the $r,\eta$-plane that can be combined with $S^1_{\chi},S^2$ to form a four cycle which measures flux $N$ in the uplifted solution.  \textbf{Right:} A generic solution with many kinks in the line charge. There are more choices of four cycles that can be used to count the number of fivebranes creating each kink. }
\label{FluxInterceptFigure}
\end{figure}

To do this, we first note that at $r=0$ the $\chi$ circle shrinks to zero size and at $\eta=0$ the $S^2$ shrinks. This means that we can form a closed four-cycle by considering the $\chi$ circle, the $S^2$ and an arc in the $r,\eta$-plane which intercepts the $\eta$-axis near a region of constant slope (see Figure \ref{FluxInterceptFigure}). Note that at this point, $\dot{V}'$ is the constant slope of this segment and the $C_3$ field (\ref{U1SymSolution}) takes the following form:

\begin{align}\label{U1SymSolutionC3}
  C_3 &\approx 2\kappa_{11} \left[
  (-\dot{V}+\eta\dot{V}')d\chi + \left({\dot{V}\dot{V}'\over\Tilde{\Delta}}-\eta\right)(d\beta + \dot{V}'d\chi)
  \right]\wedge d\Omega_{S_2}.
\end{align}

We may now find the flux of $F_4$ on this cycle by using (\ref{U1SymSolutionC3}) to calculate the difference between $C_3$ at the two endpoints of the arc. If $\lambda (\eta)$ takes the form $s_i \eta + \lambda_i$ along the segment under consideration, we find that $Q_4 = 2\lambda_i$. We can therefore interpret these intercepts as counting the number of fivebranes at each location where the slope changes. 

\subsection{Electrostatic solution for uplifted solution}

Using the map of our original coordinates $y,\theta$  to LLM coordinates $\xi,\rho$, we can express the electrostatic variables in terms of $y, \theta$. The first relation in (\ref{changev}) gives
\begin{align}\label{rdefine}
r=N\sqrt{y^2-y+q_1}\sin\theta.
\end{align}
The exact differential $d\eta$  (\ref{detadiff}) expressed in terms of the $y, \theta$  variables is given by
 \begin{align}
 d\eta =N({1\over 2}-y) \sin\theta d\theta+ N\cos \theta dy
 \end{align}
 which can be integrated to give the map from $y,\theta$ to $\eta, \xi$
\begin{align}
\eta = N(y-{1\over 2}) \cos\theta, \quad \quad  \xi= N y \cos\theta.
\end{align}
It follows from  (\ref{rdefine}) that $r=0$ corresponds to either $y=y_c$   or $\theta=0$. Plugging this relation into (\ref{lambdadef})  determines the line charge
\begin{align}
    \lambda(\eta)=\left\{ \begin{array}{cc}
     {y_c\over y_c-{1\over2}} \eta &  0< \eta< N(y_c-{1\over 2} )\\
       \eta+{N\over 2}  & \eta>N(y_c-{1\over 2}).
    \end{array}\right.
\end{align}
Using the relation of the  charge $q_1$ (\ref{q1val}) and $y_c$  (\ref{ycloc}) for a $\mathbb{R}^4/\mathbb{Z}_k$ conical singularity  with $n=2,3,\cdots$ then gives

\begin{align}
    \lambda(\eta)=\left\{ \begin{array}{cc}
     (k+1)\eta &  0< \eta<{N\over 2k}  \\
      \eta+{N\over 2}  & \eta>{N\over 2k}.
    \end{array}\right. 
\end{align}
We note that $k=1$ corresponds to $q_1=0$ and hence the $AdS_7\times S^4$ vacuum. We have $\lambda(\eta=0)=N/2$ which corresponds to a four form flux of $N$.  Note that at $y=y_c=N/(2k)$ the slope of the line charge density $\lambda(\eta)$ changes from $1$ to $k+1$.

\subsection{Generalization of electrostatic solution}

We showed in the previous section that the uplifted defect solution corresponds to a specific line charge in electrostatic formulation. In  \cite{Gaiotto:2009gz} general conditions on the line charge distribution which are imposed by charge conservation and regularity, which we will briefly review.

First, we previously remarked upon the relationship between the $F_4$ flux and the intercepts of the line charge. Imposing charge quantization therefore quantizes these intercepts. Next, in order to find constraints on the slopes, we zoom into a region of constant charge density near $r = 0$ where (\ref{U1SymSolution}) takes the form

\begin{align}\label{metricnearr0}
ds^2 &\approx \kappa_{11}^{2/3}\left({\dot{V}\Tilde{\Delta} \over 2V''}\right)^{1/3}\left(4d_{AdS_5}^2 + {2V''\dot{V}\over \Tilde{\Delta}}ds_{S^2}^2+{2V''\over\dot{V}}(dr^2+r^2d\chi^2 +d\eta^2) + {4\over \Tilde{\Delta}}(d\beta + \dot{V}'d\chi)^2\right), \nonumber \\
\Tilde{\Delta} &\approx 2\dot{V}V'' + (\dot{V}')^2 .
\end{align}

As we mentioned previously, at $r=0$ the $\chi$-circle is shrinking however the circle $\beta + \dot{V}'\chi$ is not and so we can use it to define a new periodic coordinate provided that $\dot{V}'$ takes integer values there. Since $\dot{V}'(r=0,\eta)$ is just the slope of the constant line segment, we find that regularity imposes our next quantization condition on $\lambda(\eta)$. 

There are further constraints on the changes in slope which we can deduce by zooming in on the region $\eta = \eta_i$ where two slopes meet. Here $V''$ has a delta function source which means that 
\begin{align}
V''\approx {k \over 2} {1\over \sqrt{r^2 +(\eta -\eta_i)^2}}
\end{align}
where $k$ is the change in slope. When we insert this into the metric (\ref{metricnearr0}), we find that the $r,\eta$ and circle directions give us a space that is locally $\mathbb{R}^4/\mathbb{Z}_k$. Imposing regularity, therefore, quantizes the change in slope so that it takes on (positive) integer values. It can be shown that these $A_{k-1}$ $(k>1)$ singularities give rise to non-abelian gauge fields in $AdS_5$ corresponding to global symmetries  \cite{Gaiotto:2009gz}. 

Finally, we can consider the geometry of our solution along the $\eta$-axis between two points $\eta_i,\eta_{i+1}$ at which the slope of $\lambda$ changes. Along any of these segments, we can form a closed four cycle by considering the segment $[\eta_i,\eta_{i+1}]$, the $S^2,$ and the circle $\beta + \dot{V}'\chi$. Notice that at either endpoint, $V''$ and hence $\Tilde{\Delta}$ blows up causing the circle to shrink. One can then use (\ref{U1SymSolutionC3}) to find that the flux of $F_4$ on this cycle is $\eta_{i+1}-\eta_{i}$. This can also be done for the first segment $[0,\eta_1]$ since the $S^2$ (but not the circle) shrinks at $\eta = 0$. Flux quantization thus constrains all $\eta_i$'s to take on integer values. 

In summary, we find that charge distributions give rise to regular (or with $A_k$ singularities) solutions provided that they are piecewise linear, have (decreasing) integer slopes and half-integer intercepts, and change slope only at integer values of $\eta$. Putting these together, we can write a multi-kink generalization of the uplifted flux $N$ solution:

\begin{align}\label{mkinklambda}
\lambda (\eta) = \begin{cases} 
      s_1 \eta & \eta \in [0,\eta_1] \\
      s_2 \eta + \lambda_2 & \eta \in [\eta_1,\eta_2]\\
      s_3 \eta + \lambda_3 & \eta \in [\eta_2,\eta_3]\\
      ... & ... \\
      \eta + N/2 & \eta \in [\eta_{n_{\text{kink}}},\infty).
   \end{cases}
\end{align}

Note that the continuity of $\lambda(\eta)$ alone is enough to determine the $\eta_i$'s in terms of the slope and intercept data. That is, $\eta_i = (\lambda_{i+1}-\lambda_i)/(s_{i}-s_{i+1})$ which can be written in terms of the slope changes $k_i\in\mathbb{Z}$ and the number of fivebranes creating the punctures $N_i$ to give $\eta_i = N_i/2k_i \in \mathbb{Z}$. Substituting these into (\ref{mkinklambda}) gives

\begin{align}\label{linecharge}
\lambda (\eta) = \begin{cases} 
      \left(1 + \sum_{i=1}^{n_{\text{kink}}} k_i\right) \eta & \eta \in [0,N_1/2k_1] \\
      \left(1 + \sum_{i=2}^{n_{\text{kink}}} k_i\right) \eta + N_1/2 & \eta \in [N_1/2k_1,N_2/2k_2]\\
      \left(1 + \sum_{i=3}^{n_{\text{kink}}} k_i\right) \eta + (N_1+N_2)/2 & \eta \in [N_2/2k_2,N_3/2k_3]\\
      ... & ... \\
      \eta + N/2 & \eta \in [N_{n_{\text{kink}}}/2k_{{n_{\text{kink}}}},\infty)  .
   \end{cases}
\end{align}
where $N = \sum_{i=1}^{n_{\text{kinks}}} N_i$ is the total $F_4$ flux. One can plug this general solution into (\ref{U1SymSolution}) and find that it produces the same asymptotic $AdS_7\times S^4$ region (\ref{asymptoticAdS7metric}) as the original uplifted solution. 

\section{Holographic observables}

The supergravity solutions presented in the previous section  can be used to calculate holographic observables. Examples of such observables are the entanglement entropy of a surface around the defect and the on-shell action. Due to the infinite volume of the asymptotic $AdS_7\times S^4$ region, the holographic observables are divergent and have to be regularized.   We can define a general cutoff surface 
\begin{align}\label{cutsurf}
    \eta(\epsilon,\theta)  = y_c(\epsilon, \theta) \sin \theta , \quad r(\epsilon, \theta) =   y_c(\epsilon, \theta) \cos\theta,
\end{align}
where 
\begin{align}\label{cutoffy}
    y_c(\epsilon, \theta) ={1\over \epsilon} + f_0(\theta) + f_1(\theta) \epsilon + f_2(\theta)\epsilon^2 
\end{align}
and $f_i(\theta)$ are arbitrary bounded functions of the angle $\theta \in [0,{\pi\over 2}]$.  The observables which we will consider here turn out to be integrals of   total derivatives and become integrals over the boundary of the integration regions which is given by the integral along the $\eta$ and $r$ axis as well as the cutoff surface at large $y_c$.  The cutoff of the integral along the $\eta$ and $r$ axis is given by setting $\theta=\pi/2$ and $\theta=0$  in  (\ref{cutsurf}) respectively.
The simplest choice for a cutoff surface would be given by setting all $f_i=0$ which corresponds to a circular quarter arc in the $\eta,r$ plane whose radius will go to infinity as $\epsilon\to 0$.  

\begin{figure}[t]
    \centering
    \includegraphics[scale=.6]{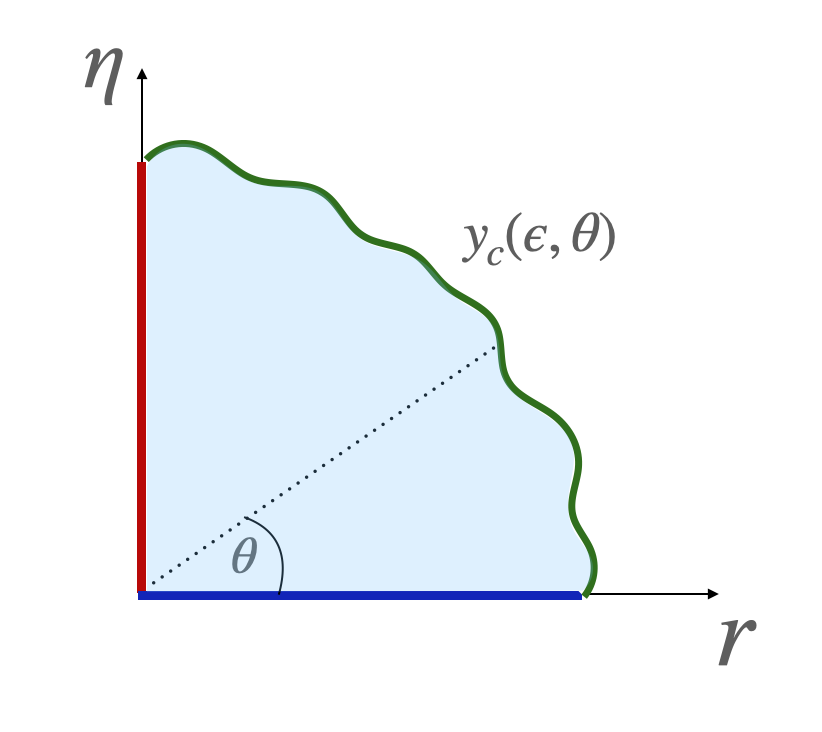}
    \caption{Integration region in the $\eta,r$-plane. We consider observables which reduce to integrals over the boundary comprised of the $\eta$-axis, $r$-axis, and a generic $\theta$-dependent cutoff surface.}
    \label{regionofintegration}
\end{figure}

In order to obtain finite results we use vacuum subtraction, i.e. we subtract the regularized result by the result for the $AdS_7\times S^4$ vacuum using the same cutoff surface. We use this prescription since a full set of covariant counterterms is  not known for the eleven-dimensional supergravity and the standard method of holographic renormalization \cite{deHaro:2000vlm,Bianchi:2001kw} which can be used for $AdS$ solutions of gauged supergravities  in lower dimensions is not available.  

The contributions from the cutoff surface can all be expressed in terms of moments of the large $y_c$ expansion of derivatives of the potential $\dot V, V''$   (\ref{vpot})

\begin{align}
    \dot V &= y_c \sin\theta+ m_1 \sin\theta- m_3 {\cos^2\theta \sin\theta\over 2 y_c^2}+ \mathcal{O}\left({1\over y_c^4}\right),\nonumber \\
    V''&= m_1 {\sin\theta\over y_c^2}-m_3{\sin\theta (1+ 5 \cos 2\theta) \over 4y_c^4}  +\mathcal{O}\left({1\over y_c^6}\right).
\end{align}
The moments $m_1$ and $m_3$ can be expressed in terms of line charge (\ref{linecharge})
\begin{align}
    m_1 &= \sum_{i=1}^{n_{\text{kinks}}}(s_i-s_{i+1})\eta_i = {1\over 2}\sum_{i=1}^{n_{\text{kinks}}} N_i = {N\over 2},\\ 
    m_3&=\sum_{i=1}^{n_{\text{kinks}}}(s_i-s_{i+1})\eta^3_i ={1\over 8}\sum_{i=1}^{n_{\text{kinks}}} {N_i^3\over k_i^2},
\end{align}
where $s_{n_{\text{kinks}+1}} = 1$. When it is unambiguous, we will just write $m_i$ but when we refer to a particular or multiple solutions at once (as in the case of vacuum subtraction), we will denote the moments with a superscript, e.g. $m_i^{(n_{\text{kinks}})}$ or $m_i^{(vac)}$.

\subsection{Central charge}

It was argued in \cite{Gauntlett:2006ai,CKKL} that the holographic  dual of the $a$ central charge of a $d=4$ SCFT 
coming from the $11$ dimensional metric
\begin{align}
    ds^2_{11} &= \kappa_{11}^{2/3} \left( {\dot V \Delta \over 2 V''}\right)^{1/3} [4 ds^2_{\AdS_5} + ds^2_{M_6}]
\end{align}
is give by the  following expression
\begin{align}
    a = \frac{2^5 \pi^3 \kappa_{11}^3}{(2 \pi \ell_p)^9} \int_{M_6} \left( {\dot V \Delta \over 2 V''}\right)^{3/2} d\Omega_{M_6},
\end{align}
where $\ell_p$ is the $11$ dimensional Planck length and $d\Omega_{M_6}$ is the volume form of $ds^2_{M_6}$. For holographic duals of $d=4,\;\cN=2$ SCFTs the six dimensional space is compact and one obtains a finite result for the integral. As discussed above, for the defect solutions the integral will be taken over a non-compact space and will be divergent.
\begin{align}
    d\Omega_{M_6} = \frac{8 \sqrt{2} r (V'')^{5/2}}{\dot V^{1/2} \Delta^{3/2}} d\Omega_{S_2} \wedge d\eta \wedge dr \wedge d \chi \wedge d \beta.
\end{align}
The central charge is therefore equal to
\begin{align}
    a &= \frac{2^{7} \pi^3 \kappa_{11}^3}{(2 \pi \ell_p)^9} \int r \dot{V} V'' d\Omega_{S_2} \wedge d\eta \wedge dr \wedge d \chi \wedge d \beta.
\end{align}
We can now use the cylindrical Laplace equation (\ref{cyllaplace}) to write $r \dot{V} V'' = - \partial_r (\dot{V}^2)/2$ and the fact that $\chi$ and $\beta$ are $2 \pi$ periodic, as well as $\kappa_{11}= {\pi\over 2} \ell_p^3$ to write the central charge as
\begin{align}
    a &= {1\over 4} \int -\partial_r (\dot{V}^2) dr \wedge d\eta \nonumber \\
    &= {1\over 4} \int_0^{y_c(\epsilon, \pi/2)}  d\eta \; \lambda(\eta)^2 - {1\over 4}\int_{\theta=0}^{\theta=\pi/2} (\dot V)^2 d\big( y_c(\epsilon,\theta)\sin\theta\big) \\
    &= {1\over 4}\int_0^{\eta_{n_{\text{kink}}}} d\eta \; \lambda(\eta)^2 + {1\over 4}\int_{\eta_{n_{\text{kink}}}}^{y_c(\epsilon,\pi/2)} d\eta \; (\eta + m_1)^2 - {1\over 4}\int_{\theta=0}^{\theta=\pi/2} (\dot V)^2d\big( y_c(\epsilon,\theta)\sin\theta\big),\nonumber
\end{align}
where we obtain the final line by noticing that $\lambda(\eta)$ has a universal form in the region beyond the final kink $\eta_{n_{\text{kink}}}$. Notice above that the first integral in the third line is finite. Inserting the generic cutoff surface (\ref{cutoffy}) into this expression and integrating over $\theta$ gives us following:
\begin{align}\label{centralchargeexpanded}
a &= {m_1/3+f_0(\pi/2) \over 4\epsilon^2}+{2m_1^2/3 + 2m_1f_0(\pi/2) +f_0(\pi/2)^2 + f_1(\pi/2) \over 4\epsilon} \nonumber\\
&+{1\over 4}\int_0^{\eta_{n_{\text{kink}}}} d\eta \; \lambda(\eta)^2 \nonumber + {1\over 60}\Big[2m_3 + 15m_1^2(-\eta_{n_{\text{kink}}}+f_0(\pi/2)) +15m_1(-\eta_{n_{\text{kink}}}^2+f_0(\pi/2)^2)\nonumber\\
&+ 5(-\eta_{n_{\text{kink}}}^3+f_0(\pi/2)^3)  +30(m_1+f_0(\pi/2))f_1(\pi/2) + 15f_2(\pi/2)\Big]\nonumber\\
&+ \int_{0}^{\pi/2}I_{m_1,f_i}(\epsilon,\theta)d\theta + \mathcal{O}(\epsilon),
\end{align}
where in the final line, $I_{m_1,f_i}(\epsilon,\theta)$ is an expression that depends on the cutoff surface functions $f_i$ but only $m_1$ and therefore, since this is the same for all solutions, it will be eliminated by subtracting the contribution from the $AdS_7\times S^4$ vacuum solution:
\begin{align}
a_{(vac)} &= {m_1/3+f_0(\pi/2) \over 4\epsilon^2}+{2m_1^2/3 + 2m_1f_0(\pi/2) +f_0(\pi/2)^2 + f_1(\pi/2) \over 4\epsilon} \nonumber\\
&+ {1\over 60}\Big[-13m_1^3 + 15m_1^2f_0(\pi/2) +15m_1f_0(\pi/2)^2\nonumber\\
&+ 5f_0(\pi/2)^3 +30(m_1+f_0(\pi/2))f_1(\pi/2) + 15f_2(\pi/2)\Big]\nonumber\\
&+ \int_{0}^{\pi/2}I_{m_1,f_i}(\epsilon,\theta)d\theta + \mathcal{O}(\epsilon).
\end{align}
This expression can be obtained from (\ref{centralchargeexpanded}) by noticing that $m^{(vac)}_3 = m_1^3$ and $\eta^{(vac)}_{n_{\text{kink}}} = \eta^{(vac)}_1 = m_1$. All of the divergent terms depend only on $m_1$. Furthermore, the cutoff surface functions, $f_i$, only appear in the finite term with $m_1$ (and no higher moments) so after vacuum subtraction we will be left with something finite and independent of the choice of cutoff:
\begin{align}
a-a_{(vac)} &= {1\over 4}\int_0^{\eta_{n_{\text{kink}}}} d\eta \; \lambda(\eta)^2 + {1\over60} (13m_1^3+2m_3 - 15m_1^2\eta_{n_{\text{kink}}} - 15m_1 \eta_{n_{\text{kink}}}^2-5\eta_{n_{\text{kink}}}^3).
\end{align}

It is useful to rewrite these expressions in terms of the more physical parameters $k_i$ and $N_i$. For one and two kinks these become
\begin{align}
a_{\text{(2)}}-a_{(vac)} &= {(-3+k_1(-10+13k_1)-5k_2)N_1^3 \over 480 k_1^2} + {39N_1^2N_2\over480} \nonumber\\
&+ {3(-5+13k_2)N_1N_2^2 \over 480 k_2}+{(-1+k_2)(3+13k_2)N_2^3\over 480k_2^2} 
\end{align}
and 
\begin{align}
a_{\text{(1)}}-a_{(vac)} = {(k-1)(3+13k)N^3\over 480 k^2}.
\end{align}

\subsection{On-shell action}
For holographic defect solutions, among the simplest observables is the vacuum subtracted on-shell action which gives the defect partition function in the semi-classical approximation.  
Other observables, which we will not discuss here, include one-point functions of bulk operators in the presence of the defect or the entanglement entropy in the presence of the defect. 

The action of eleven dimensional supergravity is given by
\begin{align}
S= {1\over 2 k_{11}^2}\int_{\cal M}  \sqrt{-g} \Big(R-{1\over 48} F_{\mu\nu\rho\lambda} F^{\mu\nu\rho\lambda}\Big)+ {1\over 2 k_{11}^2} \int_{\partial {\cal M}} \sqrt{h}\;  2K + S_{CS}.
\end{align}
Here $S_{CS}$ is the Chern-Simons term which vanishes for the LLM solutions and is dropped in the following. The second term is the Gibbons-Hawking term which is needed for a good variational principle for spacetimes with boundary.  Here $h_{ab}$ is the induced metric on the boundary and $K$ is  the trace of the second fundamental form $K_{\mu\nu}= -{1\over 2}(\nabla_\mu n_\nu+ \nabla_\nu n_\mu) $ where $n_\mu$ is the outward pointing normal vector to the boundary $\partial {\cal M}$. Using the equations of motion for the metric and the three form potential, it is easy to show that the bulk part of the on-shell action is a total derivative and the total action is given by a boundary term
\begin{align}\label{onshell}
    S_{\rm on\; shell}&= {1 \over 2  k_{11}^2}\int_{\partial {\cal M}} \left(-{1\over 3} \right)C_3\wedge *F_4 +  {1\over 2 k_{11}^2} \int_{\partial {\cal M}} \sqrt{h}\;  2K.
\end{align}
Presently, we will compute this for the simple cutoff ($f_i = 0$ for all $i$) and later comment on generic cutoff-dependence. To start, we notice that the boundary region $\eta=0$ gives no contribution since here the $S^2$ shrinks to zero volume. The contribution coming from the cutoff surface has a universal form for all solutions in terms of moments $m_1$ and $m_3$:
\begin{align}\label{Actionlargey}
S_{bulk,cutoff} = {Vol(AdS_5)Vol(S^2)\over 2 k_{11}^2}\left(-{64(-2m_1+5m_3)\over 15m_1 \epsilon} -{128(m_1^3+2m_3)\over 15}\right)
\end{align}
and 
\begin{align}
S_{GH,cutoff} = {Vol(AdS_5)Vol(S^2)\over 2 k_{11}^2}\left({128\over \epsilon^3} + {512m_1\over 3\epsilon^2} + {512m_1^2\over 15\epsilon} + {128(m_1^3-3m_3)\over 15}\right).
\end{align}
This is not unexpected since we take this boundary to be at a distance far away from the region where the slopes of $\lambda(\eta)$ are changing ($y_c(\epsilon,\theta) \gg \eta_{n_{kink}}$). 

The final contribution comes from the region along the $\eta$-axis. Since this involves an integral over $0<\eta<y_c(\epsilon,\pi/2)$, it will be sensitive to line charge data beyond just the moments. These integrals quickly become unwieldy for more complicated $\lambda(\eta)$ so in lieu of a generic expression, we can write down the answer for $n_{kink}=2$ from which the $n_{kink}=1$ case can be easily derived by setting $N_1\to 0$ and $N_2\to N$. We have that
\begin{align}\label{Actionr0}
S^{(2)}_{bulk,r=0} &= {Vol(AdS_5)Vol(S^2)\over 2 k_{11}^2}\left(-{64\over 3\epsilon^3} - {64m_1\over \epsilon^2} - {64(4m_1^3-m^{(2)}_3)\over 3m_1\epsilon} +S^{(2),finite}_{bulk,r=0}\right), \\
S^{(2),finite}_{bulk,r=0} &= {64\over 3}\left[(1+4s_1-2s_2)(s_1-s_2)\eta_1^3 + 6(s_1-s_2)(s_2-1)\eta_1\eta_2^2 + (s_2-1)(4s_2-1)\eta_2^2\right], \nonumber\\
S^{(2)}_{GH,r=0} &= {Vol(AdS_5)Vol(S^2)\over 2 k_{11}^2}\left({128\over 2\epsilon^3} + {128m_1\over\epsilon^2} + {128m_1^2 \over \epsilon} +S^{(2),finite}_{GH,r=0}\right),\\
S^{(2),finite}_{GH,r=0} &= {-128\over3}\left[m^{(2)}_3(2s_1-s_2) + (s_2-1)(3m_1-2(s_1-1))\eta_2^2\right].\nonumber
\end{align}
One can quickly inspect that the $\epsilon^{-3}$ and $\epsilon^{-2}$ divergences only depend on $m_1$ and so will cancel once we subtract the vacuum contribution. There are $m_3$'s which appear in the $\epsilon^{-1}$ divergent term, however they cancel between  (\ref{Actionlargey}) and (\ref{Actionr0}). Combining all of the terms, we obtain the following
\begin{align}
S^{(2)}_{\text{on shell}} ={2\pi Vol(AdS_5)\over k_{11}^2}\left({448\over 3\epsilon^3} + {704m_1\over 3\epsilon^2} + {256m_1^2\over 3\epsilon} - {64\over 3}m^{(2)}_3\right)
\end{align}
and after subtracting the $AdS_7\times S^4$ vacuum, we are left with
\begin{align}
S^{(2)}_{\text{on shell}} - S^{(vac)}_{\text{on shell}}&={-2\pi Vol(AdS_5)\over  k_{11}^2}{64\over 3}(m^{(2)}_3-m_3^{(vac)}) \nonumber\\
&= {16\pi Vol(AdS_5)\over  3k_{11}^2}\left((N_1+N_2)^3-{N_1^3\over k_1^2}+{N_2^3\over k_2^2} \right).
\end{align}
From this, we can set $N_1= 0$, $N_2 = N$ and $k_2 = k$ to obtain the expression for one kink:
\begin{align}\label{finite-on-shell}
S^{(1)}_{\text{on shell}} - S^{(vac)}_{\text{on shell}}
&= {16Vol(AdS_5)\over 3 k_{11}^2}N^3\left( 1 -{1\over k^2} \right)\nonumber\\
&={-2\pi Vol(AdS_5)\over  k_{11}^2}{64\over 3}(m^{(1)}_3-m_3^{(vac)}).
\end{align}
The terms with ${1\over\epsilon^{2n}} $ divergences  cancel out of the vacuum  subtracted on shell action. However, the result still has a divergence due to the infinite volume of $AdS_5$. For a more complete treatment one should introduce a Fefferman-Graham like cutoff which regularizes all divergences, see e.g. \cite{Estes:2014hka,Gutperle:2016gfe} for discussions  of such cutoffs in other holographic defect theories.

Another possible related feature of the vacuum subtracted on shell action is that the detailed form of finite terms depend on the choice of the cutoff surface.  This is analogous to the possibility of finite counter terms in a covariant regularization procedure in lower dimensional supergravity. Such ambiguities can often be fixed by demanding the finite counter terms preserve supersymmetry, but how this implemented in the vacuum subtraction is not clear to us at this moment. While the results for a simple cutoff we have presented in this section are compellingly simple, it is not clear at the moment whether they are unambiguous. 

\subsection{Defects in the  dual SCFT}
A co-dimension two  conformal defect  in a six dimensional CFT preserves a $SO(4,2)\times SO(2)$ subgroup of  $SO(6,2)$.  For the $d=6,N=(2,0)$ SCFT which are dual to the $AdS_7\times S^4$ vacua of M-theory, the superconformal symmetry is $OSp(8^*|2)$ and a half BPS-defect co-dimension two  defect that our supergravity solutions preserve a $SU(2,2|2)$ defect conformal sub algebra. See   \cite{DHoker:2008wvd} for a classification of conformal sub algebras which correspond to half-BPS defects of maximally supersymmetric SCFTs.  The general analysis for less supersymmetry and arbitrary co-dimension  has not been performed to our knowledge, see however \cite{Agmon:2020pde} for a complete analysis for conformal line defects in SCFTs.  

It is a challenge to construct explicit duals on the CFT side of the supergravity solutions describing defects constructed in this paper, since the $d=6,N=(2,0)$ SCFT  does not have a Lagrangian formulation.  It is often useful to construct a defect in a simpler theory and we do this using in appendix \ref{appendixa} for the theory of a free six dimensional $N=2$ hypermultiplet. The field theory  defect solution is given by a nontrivial profile of for two of  the five scalars in the tensor multiplet in the two directions transverse to the defect. This  construction is analogous to the construction of surface defects in $d=4,N=4$ SYM  due to Gukov and Witten \cite{Gukov:2006jk}. The free tensor multiplet provides only a simple model for  the ``center of mass" degrees of freedom and the construction of the defect solution for the strongly coupled interacting $d=6, \; \cN=(2,0)$ is a much harder problem.

As mentioned above the  defect theory has $SU(2,2|2)$ superconformal symmetry which is the same as $N=2,d=4$ SCFTs. This is no surprise since our supergravity solutions are closely related to LLM and Gaiotto-Maldacena solutions as discussed above, which can be interpreted as coming from compatifications of  M5-branes on compact Riemann surfaces with punctures. 
It is interesting to contrast these holographic solutions with the ones used to describe $d=4,\; \cN=2$ SCFTs  \cite{Gaiotto:2009gz,Reid-Edwards:2010vpm,Aharony:2012tz,Petropoulos:2013vya}   as well as more recent ones constructing duals of Argyres-Douglas theories \cite{Bah:2021mzw,Bah:2021hei,CKKL}. In the former, the $\eta,r$ is 
compact and will be related to Maldacena-Nunez \cite{Maldacena:2000mw} solutions and  class S $\cN=2,d=4$ theory \cite{Gaiotto:2009we}  coming from compactifying a $d=6, \; \cN=(2,0)$ theory on a Riemann surface with (regular) punctures. In the latter, one considers a disk in the $\eta,r$  
plane with 5-brane source smeared on the boundary of the disk. This behavior is to be contrasted to   our solutions  where the $\eta,r$ space is non-compact and the 
solutions are asymptotically 
$AdS_7\times S^4$  in the limit where $\eta,r$ go to infinity. Hence the supergravity solutions are holographically dual to    co-dimension 2 defects in $d=6, \; \cN=(2,0)$ SCFTs.  Most   solutions are singular with singularities corresponding to a finite number of  regular punctures, associated with the kinks in the linear charge density.  It is however possible to  construct solutions where the slope of the kinks only changes by one and hence they  are be completely regular. 

Since the superconformal 
symmetry preserved by the defect is the same as the one of 
$d=4, \; \cN=2$ SCFTs it is natural that these SCFTs describe the defect degrees of freedom. For solutions with regular punctures it is likely that the defect theories can be related to the generalized quiver theories of \cite{Gaiotto:2009we}.  It is an open questions how to interpret the completely regular solutions.
The calculation of some holographic observables given in this paper is a first step in checking any identification of defect theories. It may be possible to check the identification by matching holographic calculation with calculations on the field theory side using localization. These interesting questions are currently under investigation.

\section{Discussion}

In this paper we constructed solutions of eleven dimensional supergravity, which are holographic duals of co-dimension two defects in six dimensional SCFTs. The solutions preserve sixteen of the thirty two supersymmetries. 

While it is possible to construct completely regular quarter-BPS solutions which carry two nonzero charges, the seven dimensional half-BPS solution with only one nonzero charge turned on suffers from a conical singularity in the bulk. Upon uplifting to eleven dimensions we showed that the singularity is also present in eleven dimensions.  The uplift allows us to identify this type of singularity with a regular puncture which is locally $\mathbb{R}^4/\mathbb{Z}_k$ and was discussed already in the original paper of Gaiotto and Maldacena  \cite{Gaiotto:2009gz} that constructs holographic duals of $d=4,\; \cN=2$ SCFTs.

One of the main results in the present paper is to use the electrostatic formulation of the  LLM solution to construct new defect  solutions based on more general linear charge densities. It is possible to obey all the conditions that   charge quantization  and periodicity of the angular coordinates impose.  The generic solutions have singularities corresponding to a finite number of  regular punctures, associated with the kinks in the linear charge density.  It is however possible to  construct solutions which can be completely regular. 

We note that the  electromagnetic formulation involves an approximation where we consider a rotationally symmetric distribution of sources  for the Toda equation and smear them. It would be interesting to consider solutions of the Toda equation corresponding to co-dimension two defect solutions. This would involve placing line sources in the  three dimensional  half space spanned by $\xi, x_1,x_2$. The holographic defects  would correspond to solutions where this  space is non-compact and the large $\xi, x_i$ limit corresponds to the asymptotic $AdS_7\times S^4$ region. The nonlinear nature of the Toda equation which determines the solution makes the construction of such solutions very challenging, however.  It would also be interesting to find generalizations of the uplifts of  the quarter-BPS defect solutions which are completely regular already in 7 dimensions. Since no general ``bubbling" solution \`a la LLM exists for eight instead of sixteen preserved supersymmetries, this also is a question which we will leave for the future. 

\bigskip

\section*{Acknowledgements}

The authors would like to thank T. Dumitrescu for useful conversations.
The work of M. G. was supported, in part, by the National Science Foundation under grant PHY-2209700. The authors are grateful to the Mani L. Bhaumik Institute for Theoretical Physics for support.

\newpage

\appendix

\section{Defects for the $d=6, \; \cN=(2,0)$ tensor multiplet}\label{appendixa}

In this appendix we construct a conformal co-dimension two defect for the free $d=6, \; \cN=(2,0)$ tensor multiplet. The field content of the multiplet is a rank 2 antisymmetric tensor field $B_{\mu\nu}$ with self-dual field strength $H_{\mu\nu\rho}$, five scalars $\Phi_i, i=1,\cdots,5$  which transform as a ${\bf 5} $ under the $SO(5)$ R-symmetry and four  symplectic Majorana-Weyl spinors $\psi^a, a=1,\cdots,4$ which transform as ${\bf 4}$ under the $USp(4)\equiv SO(5)$.

The super(conformal) symmetry transformations are given by \cite{Claus:1997cq}

\begin{align}\label{susytensor}
\delta\psi &= {1\over 2} \gamma^\mu \partial_\mu \phi_i \Gamma^i\epsilon-{1\over 6} H_{\mu\nu\rho}\gamma^{\mu\nu\rho}\epsilon+ 2 \phi_i \Gamma^i \eta_0,\nonumber\\
\delta \phi^i &= -2\bar \epsilon   (\Gamma^i)   \psi,\nonumber\\
\delta B_{\mu\nu} &= -2 \bar \epsilon \gamma_{\mu\nu}\lambda.
\end{align}
Here $\Gamma^i, i=1,2,\cdots,5$ are $SO(5)$ gamma matrices and $\gamma^\mu$ are six dimensions gamma-matrices. The spinors are contracted using the symplectic metric $\Omega_{ab}$. The supersymmetry transformation parameter   $\epsilon$ is given by
\begin{align}
\epsilon = \epsilon_0+ \gamma_\mu x^\mu \eta_0,
\end{align}
where $\epsilon_0$ is a left handed constant symplectic Majorana spinor parameterizing the Poincare supersymmetries annd   $\eta_0$ is a constant right handed   symplectic Majorana spinor, parameterizing the superconformal transformations.  We are constructing a   co-dimension two defect in this six dimensional theory, which preserves some part of the superconformal symmetry. The simplest set-up is to consider a flat defect with a four dimensional world-volume directions, on which all  fields  do not depend. The two directions transverse to the defect  are spanned by  $x^1,x^2$ and we choose the defect to be located at $x^1=x^2=0$. From the symmetries we can deduce  that the antisymmetric tensor field is vanishing and hence only the scalars  are turned on. It is useful to introduce complex coordinates $z=x^1+ i x^2$ and gamma matrices
\begin{align}
\gamma^z={1\over \sqrt{2} }\Big( \gamma^1+ i \gamma^2\Big), \quad \quad \gamma^{\bar z}={1\over \sqrt{2}} \Big( \gamma^1- i \gamma^2\Big).
\end{align}

From the supergravity solutions it follows that for a defect that preserves half the supersymmetries the $SO(5)$ R-symmetry is broken to $SU(2)$, hence we make the following ansatz for the scalar fields.  The following  complex combination of the scalar fields is nontrivial

\begin{align}\label{tensorscal}
\phi_\omega = {1\over \sqrt{2}} \Big(\phi^1+i \phi^2\Big)={\alpha+ i\beta\over z}.
\end{align}
Unbroken supersymmetries satisfy $\delta\psi=0$, it is easy to see that the supersymmetry transformation rules (\ref{susytensor})  lead to the condition on the Poincare supersymmetry 
\begin{align}\label{tensorproj}
\gamma^a \Gamma^\omega \epsilon_0 =0\quad   \Leftrightarrow  \quad \gamma^{12}\Gamma^{12}\epsilon_0=\epsilon_0.
\end{align}
The second condition is a projection which implies that half the Poincare supersymmetries are preserved.
It is also easy to verify that for an $\eta_0$ satisfying the same projection condition (\ref{tensorproj}) and the $z$ dependence of the scalar (\ref{tensorscal}) half of the superconformal symmetries are preserved and hence the defect is half BPS.

For a defect preserving a quarter of the supersymmetry we have a nontrivial profile for four scalars, breaking the $SO(5)$ R-symmetry to   $U(1)\times U(1)$.
\begin{align}\label{tensorscal2}
\phi_{\omega_1} = {1\over \sqrt{2}} \Big(\phi^1+i \phi^2\Big)={\alpha_1+ i\beta_1\over z}, \quad \quad \phi_{\omega_2} = {1\over \sqrt{2}} \Big(\phi^3+i \phi^4\Big)={\alpha_2+ i\beta_2\over z},
\end{align}
which leads to two projectors 
\begin{align}
\gamma^{12}\Gamma^{12}\epsilon_0=\epsilon_0, \quad \quad  \gamma^{12}\Gamma^{34}\epsilon_0=\epsilon_0.
\end{align}
Hence a quarter of the supersymmetries are preserved (as well as a quarter of the superconformal symmetries).
The free tensor multiplet can be used to construct the $\cN=(2,0)$ superconformal current multiplet which contains the $SO(5)$ R-symmetry current and the stress tensor \cite{Bergshoeff:1999db}. The free tensor multiplet corresponds to the ``center of mass" degrees of freedom and the construction of the defect solution for the strongly coupled interacting $d=6, \; \cN=(2,0)$ theory is beyond the scope of this appendix.

\newpage

\providecommand{\href}[2]{#2}\begingroup\raggedright\endgroup

\end{document}